\documentclass[12pt,notoc]{JHEP}
\usepackage{psfrag}
\usepackage{graphics}
\usepackage{amssymb,epsfig,amsmath,euscript,array}

\makeatletter
\@addtoreset{equation}{section}
\makeatother

\newcounter{multieqs}

\newcommand{\be}{\begin{equation}}
\newcommand{\ee}{\end{equation}}

\newcommand{\bm}[1]{\mbox{\boldmath $#1$}}

\def\bd{\begin{document}}
\def\ed{\end{document}}
\def\nn{\nonumber}
\def\bea{\begin{eqnarray}}
\def\eea{\end{eqnarray}}
\let\bm=\bibitem
\let\la=\label

\newcommand{\EQ}[1]{\begin{equation} #1 \end{equation}}
\newcommand{\AL}[1]{\begin{subequations}\begin{align} #1 \end{align}\end{subequations}}
\newcommand{\SP}[1]{\begin{equation}\begin{split} #1 \end{split}\end{equation}}
\newcommand{\ALAT}[2]{\begin{subequations}\begin{alignat}{#1} #2 \end{alignat}\end{subequations}}
\def\beqa{\begin{eqnarray}}
\def\eeqa{\end{eqnarray}}
\def\beq{\begin{equation}}
\def\eeq{\end{equation}}

\def\hf{{\textstyle{1\over2}}}
\def\wbar{\bar w}
\def\mubar{\bar\mu}
\def\abar{\bar a}
\def\sigmabar{\bar\sigma}
\def\etabar{\bar\eta}
\def\zetabar{\bar\zeta}
\def\mubar{\bar\mu}
\def\nubar{\bar\nu}
\def\N{{\cal N}}
\def\sst{\scriptscriptstyle}
\def\thetabar{\bar\theta}
\def\Tr{{\rm Tr}}
\def\one{\mbox{1 \kern-.59em {\rm l}}}
 \def\Nh{\hat{N}}

\def\a{\alpha}      \def\da{{\dot\alpha}}
\def\b{\beta}       \def\db{{\dot\beta}}
\def\c{\gamma}  \def\G{\Gamma}  \def\cdt{\dot\gamma}
\def\d{\delta}  \def\D{\Delta}  \def\ddt{\dot\delta}
\def\e{\epsilon}        \def\vare{\varepsilon}
\def\f{\phi}    \def\F{\Phi}    \def\vvf{\f}
\def\h{\eta}
\def\k{\kappa}
\def\l{\lambda} \def\L{\Lambda}
\def\m{\mu} \def\n{\nu}
\def\o{\omega}
\def\p{\pi} \def\P{\Pi}
\def\r{\rho}
\def\s{\sigma}  \def\S{\Sigma}
\def\t{\tau}
\def\th{\theta} \def\Th{\Theta} \def\vth{\vartheta}
\def\X{\Xeta}
\def\z{\zeta}

\def\cA{{\cal A}} \def\cB{{\cal B}} \def\cC{{\cal C}}
\def\cD{{\cal D}} \def\cE{{\cal E}} \def\cF{{\cal F}}
\def\cG{{\cal G}} \def\cH{{\cal H}} \def\cI{{\cal I}}
\def\cJ{{\cal J}} \def\cK{{\cal K}} \def\cL{{\cal L}}
\def\cM{{\cal M}} \def\cN{{\cal N}} \def\cO{{\cal O}}
\def\cP{{\cal P}} \def\cQ{{\cal Q}} \def\cR{{\cal R}}
\def\cS{{\cal S}} \def\cT{{\cal T}} \def\cU{{\cal U}}
\def\cV{{\cal V}} \def\cW{{\cal W}} \def\cX{{\cal X}}
\def\cY{{\cal Y}} \def\cZ{{\cal Z}}

\def\ua{\underline{\alpha}}
\def\ub{\underline{\phantom{\alpha}}\!\!\!\beta}
\def\uc{\underline{\phantom{\alpha}}\!\!\!\gamma}
\def\um{\underline{\mu}}
\def\ud{\underline\delta}
\def\ue{\underline\epsilon}
\def\una{\underline a}\def\unA{\underline A}
\def\unb{\underline b}\def\unB{\underline B}
\def\unc{\underline c}\def\unC{\underline C}
\def\und{\underline d}\def\unD{\underline D}
\def\une{\underline e}\def\unE{\underline E}
\def\unf{\underline{\phantom{e}}\!\!\!\! f}\def\unF{\underline F}
\def\unm{\underline m}\def\unM{\underline M}
\def\unn{\underline n}\def\unN{\underline N}
\def\unp{\underline{\phantom{a}}\!\!\! p}\def\unP{\underline P}
\def\unq{\underline{\phantom{a}}\!\!\! q}
\def\unQ{\underline{\phantom{A}}\!\!\!\! Q}
\def\unH{\underline{H}}

\def\As {{A \hspace{-6.4pt} \slash}\;}
\def\bs {{b \hspace{-6.4pt} \slash}\;}
\def\Ds {{D \hspace{-6.4pt} \slash}\;}
\def\ds {{\del \hspace{-6.4pt} \slash}\;}
\def\ss {{\s \hspace{-6.4pt} \slash}\;}
\def\ks {{ k \hspace{-6.4pt} \slash}\;}
\def\ps {{p \hspace{-6.4pt} \slash}\;}
\def\pas {{{p_1} \hspace{-6.4pt} \slash}\;}
\def\pbs {{{p_2} \hspace{-6.4pt} \slash}\;}

\def\Fh{\hat{F}}
\def\Vh{\hat{V}}
\def\Xh{\hat{X}}
\def\ah{\hat{a}}
\def\xh{\hat{x}}
\def\yh{\hat{y}}
\def\ph{\hat{p}}
\def\xih{\hat{\xi}}

\def\psit{\tilde{\psi}}
\def\Psit{\tilde{\Psi}}
\def\tht{\tilde{\th}}

\def\At{\tilde{A}}
\def\Qt{\tilde{Q}}
\def\Rt{\tilde{R}}
\def\Nt{\tilde{N}}

\def\at{\tilde{a}}
\def\st{\tilde{s}}
\def\ft{\tilde{f}}
\def\pt{\tilde{p}}
\def\qt{\tilde{q}}
\def\vt{\tilde{v}}
\def\nt{\tilde{n}}

\def\delb{\bar{\partial}}
\def\bz{\bar{z}}
\def\bD{\bar{D}}
\def\bB{\bar{B}}

\def\bk{{\bf k}}
\def\bl{{\bf l}}
\def\bp{{\bf p}}
\def\bq{{\bf q}}
\def\br{{\bf r}}
\def\bx{{\bf x}}
\def\by{{\bf y}}
\def\bR{{\bf R}}
\def\bV{{\bf V}}

\def\d{\delta}\def\D{\Delta}\def\ddt{\dot\delta}

\def\pa{\partial} \def\del{\partial}
\def\xx{\times}
\def\uno{\mbox{1 \kern-.59em {\rm l}}}

\def\trp{^{\top}}
\def\inv{^{-1}}
\def\dag{{^{\dagger}}}
\def\pr{^{\prime}}

\def\rar{\rightarrow}
\def\lar{\leftarrow}
\def\lrar{\leftrightarrow}

\newcommand{\0}{\,\!}      
\def\one{1\!\!1\,\,}
\def\im{\imath}
\def\jm{\jmath}

\newcommand{\tr}{\mbox{tr}}
\newcommand{\slsh}[1]{/ \!\!\!\! #1}

\def\vac{|0\rangle}
\def\lvac{\langle 0|}

\def\hlf{\frac{1}{2}}
\def\ove#1{\frac{1}{#1}}

\def\Box{\square}
\def\ZZ{\mathbb{Z}}
\def\CC#1{({\bf #1})}
\def\bcomment#1{}
\def\bfhat#1{{\bf \hat{#1}}}
\def\VEV#1{\left\langle #1\right\rangle}
\def\vev#1{\langle{#1}\rangle}

\newcommand{\ex}[1]{{\rm e}^{#1}} \def\ii{{\rm i}}

\def\rr{{\rm r}} \def\rs{{\rm s}}\def\rv{{\rm v}}
\def\ri{{\rm i}}\def\rj{{\rm j}}

\newcommand{\lrbrk}[1]{\left(#1\right)}
\newcommand{\sfrac}[2]{{\textstyle\frac{#1}{#2}}}

\font\mybb=msbm10 at 12pt
\def\bb#1{\hbox{\mybb#1}}

\font\myBB=msbm10 at 18pt
\def\BB#1{\hbox{\myBB#1}}

\title{Non-MHV Tree Amplitudes in Gauge Theory}

\author{George~Georgiou, E.~W.~N.~Glover and Valentin~V.~Khoze\\
Institute of Particle Physics Phenomenology,
Department of Physics,\\
University of Durham, Durham, DH1 3LE, UK \\
E-mail: \tt george.georgiou, e.w.n.glover, valya.khoze@durham.ac.uk }

\abstract{We show how all non-MHV tree-level amplitudes in
$0\le \cN \le 4$ gauge theories can be obtained directly from the known
MHV amplitudes using the scalar graph approach of Cachazo, Svrcek and Witten.
Generic amplitudes are given by sums of inequivalent scalar diagrams with MHV vertices.
The novel feature of our method is that after the `Feynman rules' for scalar diagrams
are used, together with a particular choice of the reference spinor,
no further helicity-spinor algebra is required to convert the results into
a  numerically usable form.
Expressions for all relevant individual diagrams 
are free of 
singularities at generic phase space points, 
and amplitudes are manifestly Lorentz- (and gauge-) invariant.
To illustrate the method, we derive expressions for $n$-point amplitudes
with three negative helicities carried by fermions and/or gluons. We also write down
a supersymmetric expression based on Nair's supervertex
which gives rise to all such amplitudes in
$0\le \cN \le 4$ gauge theories.}

\preprint{ hep-th/0407027}

\begin{document}
\baselineskip 6mm

\section{Introduction}

In a recent paper Cachazo, Svrcek and Witten \cite{CSW}
proposed a remarkable new approach for calculating
tree-level scattering amplitudes of $n$ gluons.
In this approach tree amplitudes in gauge theory
are found by summing tree-level scalar diagrams.
The CSW formalism \cite{CSW} is constructed in terms of
scalar propagators, $1/q^2,$ and
tree-level
maximal helicity violating (MHV) amplitudes,
which are interpreted as new scalar vertices.
This novel diagrammatic approach follows from an earlier construction
of Witten  \cite{Witten}  which related
perturbative amplitudes of conformal $\cN=4$ supersymmetric gauge theory
to D-instanton contributions in a topological string theory
in twistor space.

The results of \cite{Witten,CSW} have been tested and further developed in gauge
theory in \cite{GK,Zhu,BBK,Kosower,CSW2},
and in string theory and supergravity in
\cite{BM,RSV,NV,W,GLMN,Siegel,Giombi,Popov,BW}.

In \cite{GK} the CSW diagrammatic approach \cite{CSW} was
extended to gauge theories with
fermions, and it was also shown that supersymmetry is not required
for the construction to work. At tree level the scalar graph formalism
works in supersymmetric and non-supersymmetric theories, including QCD.

The motivation of the present paper is to
show how non-MHV (NMHV) tree-level amplitudes in
$0\le \cN \le 4$ gauge theories can be obtained directly from the scalar graph
approach.
One of the main points we want to make is that after the
`Feynman rules' for scalar diagrams
are used, together with the off-shell continuation of helicity spinors
on internal lines,
expressions for all relevant individual diagrams are automatically free of 
unphysical
singularities  at generic phase space points, and amplitudes are manifestly Lorentz- (and gauge-) invariant.
Hence,
no further helicity-spinor algebra is required to convert the results into
an immediately usable form.

To illustrate the method, we will derive expressions for $n$-point amplitudes
with three negative helicities carried by fermions and/or gluons.
We will also write down
a supersymmetric expression which gives rise to all such amplitudes in
$0\le \cN \le 4$ gauge theories.
This compliments a very recent calculation of Kosower \cite{Kosower}
of such amplitudes in the purely gluonic case.

As in \cite{GK},
we will consider tree-level amplitudes in a generic $SU(N)$ gauge theory
with an arbitrary
finite number of colours.
$SU(N)$ is unbroken and all fields are
taken to be massless, we refer to them generically as gluons, fermions and
scalars. The gauge theory is not necessarily assumed to be supersymmetric,
i.e. the number of supercharges is $4\cN$, where $0\le \cN \le 4$.

This paper is organised as follows.
In section 2 we will use supersymmetric Ward identities to
express NMHV purely gluonic amplitudes in terms of NMHV
amplitudes with gluons and two fermions\footnote{It may be worthwhile to note
that while a gluonic non-MHV amplitude can be determined in terms of
amplitudes with fermions and gluons, the converse of this statement is not true.
Individual non-MHV amplitudes involving fermions cannot be deduced with susy
Ward identities from amplitudes with gluons only.}.
Then,
using the CSW scalar graph method for gluons \cite{CSW} and fermions \cite{GK},
in sections 3 and 4 we will derive expressions for the
NMHV amplitudes with three negative helicities involving
gluons and fermions.

Section 5 of this paper considers the scalar graph method with the single
analytic supervertex of Nair \cite{Nair}. We provide
a single formula which gives rise to all tree-level NMHV amplitudes
with three negative helicities in $0\le \cN \le 4$ supersymmetric gauge theories,
involving all possible configurations of gauge fields, fermions and scalars.
There is also no principle obstacle to continue with further iterations of the
analytic supervertex and derive formal expressions for tree amplitudes with an
arbitrary number of negative helicities. Depending on the topology of the
iteration, these expressions
would correspond to different skeleton diagrams of \cite{BBK} in
$0\le \cN \le 4$ supersymmetric gauge theories.

We end the introduction with a brief review of the spinor helicity
formalism and definitions of the MHV amplitudes.

\subsection{Amplitudes in the spinor helicity formalism}

Using colour decomposition, an $n$-point amplitude ${\cal M}_n $
can be represented as a sum of products of colour factors $T_n$
and purely kinematic partial
amplitudes $A_n$.
The latter have the colour information stripped off
and hence do not distinguish between fundamental quarks and adjoint gluinos.
The scalar graph method \cite{CSW} is used to evaluate only the
purely kinematic amplitudes $A_n.$
Full amplitudes are then determined uniquely from the kinematic part $A_n$,
and the known expressions for $T_n.$

We will first consider theories with $\cN \le 1$ supersymmetry.
Gauge theories with extended supersymmetry have a more intricate behaviour
of their amplitudes in the helicity basis
and their study will be postponed until section 5.
Theories with $\cN=4$ (or $\cN=2$) supersymmetry have $\cN$ different species of gluinos
and 6 (or 4) scalar fields. This leads to a large number of elementary MHV-like
vertices in the scalar graph formalism. This proliferation of elementary vertices
asks for a super-graph generalization of the CSW scalar graph method, which
will be outlined in section 5.

Now we concentrate on tree level partial amplitudes $A_n=A_{l+2m}$
with $l$ gluons and $2m$ fermions in the helicity basis, and all
external lines are defined to be
incoming.

In $\cN \le 1$ theory a fermion of helicity $+{1\over 2}$ is
always connected
by a fermion propagator to a helicity $-{1\over 2}$
fermion hence the number
of fermions $2m$ is always even. This statement is correct
only in theories without scalar fields. In the $\cN=4$ theory,
a pair of positive helicity fermions, $\Lambda^{1+}$, $\Lambda^{2+}$,
can be connected to another pair of positive helicity
fermions, $\Lambda^{3+}$, $\Lambda^{4+}$, by a scalar propagator.

In $\cN \le 1$ theory
a tree amplitude $A_n$ with less than two opposite helicities
vanishes\footnote{In the $\cN=1$ theory this is also correct
to all orders in the loop expansion and non-perturbatively.}
identically \cite{Grisaru}.
First nonvanishing amplitudes contain $n-2$ particles
with helicities of the same sign
\cite{PT,BG}
and are called maximal helicity violating
(MHV) amplitudes.

In the spinor helicity formalism
\cite{Berends,PT,BG} an on-shell momentum
of a massless particle, $p_\mu p^\mu=0,$ is represented as
\be
p_{a \dot a} \equiv \ p_\mu \sigma^\mu_{a \dot a}
=\ \lambda_a\tilde\lambda_{\dot a} \ ,
\ee
where $\lambda_a$ and $\tilde\lambda_{\dot a}$
are two commuting
spinors of positive and negative chirality.
Spinor inner products are defined
by\footnote{Our conventions for spinor helicities follow
\cite{Witten,CSW} and are the same as in \cite{GK}.}
\be
\langle \lambda,\lambda'\rangle = \ \epsilon_{ab}\lambda^a\lambda'{}^b
 \ , \qquad
[\tilde\lambda,\tilde\lambda'] =\ \epsilon_{\dot a\dot b}
\tilde\lambda^{\dot a}\tilde\lambda'{}^{\dot b} \ ,
\ee
and a scalar product of two null vectors,
$p_{a\dot a}=\lambda_a \tilde\lambda_{\dot a}$ and
$q_{a\dot a}=\lambda'_a\tilde\lambda'_{\dot a}$, becomes
\be \label{scprod}
p_\mu q^\mu =\ {1\over 2}
\langle\lambda,\lambda'\rangle[\tilde\lambda,\tilde\lambda'] \ .
\ee

An MHV amplitude $A_n=A_{l+2m}$
with $l$ gluons and $2m$ fermions in $\cN \le 1$ theories
exists only for $m=0,1,2$.
This is because it must have precisely $n-2$ particles with positive and
$2$ with negative helicities, and our fermions always come in pairs
with helicities $\pm {1\over 2}$.
Hence, there are
three types of MHV tree amplitudes in $\cN \le 1$ theories:
\be
\label{threecls}
A_n (g_r^-,g_s^-) \ , \quad
A_n (g_t^-,\Lambda_r^-,\Lambda_s^+)\ , \quad
A_n (\Lambda_t^-,\Lambda_s^+,\Lambda_r^-,\Lambda_q^+) \ .
\ee
Suppressing the overall
momentum conservation factor,
$ i g_{\rm YM}^{n-2} \, (2\pi)^4 \, \delta^{(4)} (\sum_{i=1}^n \lambda_{i a}
\tilde{\lambda}_{i \dot a} ),$
the MHV purely gluonic amplitude reads \cite{PT,BG}:
\be
A_n (g_r^-,g_s^-)=\
{\langle\lambda_r,\lambda_s\rangle^4\over
\prod_{i=1}^n\langle\lambda_i, \lambda_{i+1}\rangle }
\equiv \
{\vev{r~s}^4 \over \prod_{i=1}^n \vev{i~i+1}} \ ,
\label{mpng}
\ee
where $\lambda_{n+1} \equiv \lambda_1$.
The MHV amplitude with two external fermions and $n-2$ gluons is
\be
\label{ndcls}
A_n (g_t^-,\Lambda_r^-,\Lambda_s^+)= \
{\vev{t~r}^3\ \vev{t~s} \over \prod_{i=1}^n \vev{i~i+1}} \ , \quad
A_n (g_t^-,\Lambda_s^+,\Lambda_r^-)= \
-\ {\vev{t~r}^3\ \vev{t~s} \over \prod_{i=1}^n \vev{i~i+1}} \ ,
\ee
where the first expression corresponds to $r<s$ and the second to $s<r$
(and $t$ is arbitrary).
The MHV amplitudes with four fermions and $n-4$ gluons on external lines are
\be
\label{ndcls2}
A_n (\Lambda_t^-,\Lambda_s^+,\Lambda_r^-,\Lambda_q^+)
= \
{\vev{t~r}^3\ \vev{s~q} \over \prod_{i=1}^n \vev{i~i+1}} \ , \quad
A_n (\Lambda_t^-,\Lambda_r^-,\Lambda_s^+,\Lambda_q^+)
= \
-\ {\vev{t~r}^3\ \vev{s~q} \over \prod_{i=1}^n \vev{i~i+1}}
 \
\ee
The first expression in \eqref{ndcls2}
corresponds to $t<s<r<q,$ the second -- to $t<r<s<q,$
and there are other similar expressions,
obtained by further permutations of fermions, with the overall
sign determined by the ordering.

Expressions \eqref{ndcls}, \eqref{ndcls2} can be derived from supersymmetric
Ward identities \cite{Grisaru,MP,Dixon},
and we will have more to say about this in section 5.
The $\overline{\rm MHV}$ amplitude can be obtained, as always,
by exchanging helicities $+\leftrightarrow -$ and
$\vev{i~j} \leftrightarrow [i~j].$

\section{Gluonic NMHV amplitudes and the CSW method}

The formalism of CSW was developed in \cite{CSW} for calculating
purely gluonic amplitudes at tree level. In this approach
all non-MHV $n$-gluon
amplitudes (including $\overline{\rm MHV}$) are expressed
as sums of tree diagrams in an effective scalar perturbation theory.
The vertices in this theory are the MHV amplitudes \eqref{mpng},
continued off-shell as described below, and connected by scalar
propagators $1/q^2$.

It was shown in \cite{GK} that the same idea continues to work in
theories with fermions and gluons. Scattering amplitudes are
determined from scalar diagrams with three types of MHV vertices,
\eqref{mpng},\eqref{ndcls} and \eqref{ndcls2}, which are connected to each
other with scalar propagators $1/q^2$. Also, at tree level,
supersymmetry is irrelevant and
the method applies to supersymmetric and non-supersymmetric theories
\cite{GK}.

When one leg of an MHV vertex is connected by a propagator
to a leg of another MHV vertex, both legs become internal
to the diagram and have to be continued off-shell. Off-shell continuation
is defined as follows \cite{CSW}:
we pick an arbitrary spinor $\xi_{\sst\rm Ref}^{\dot a}$ and define
$\lambda_a$ for any internal line carrying momentum $q_{a\dot a}$
by
\be \label{ofsh}
\lambda_a=q_{a\dot a}\xi_{\sst \rm Ref}^{\dot a}\ .
\ee
External lines in a diagram remain on-shell, and for them
$\lambda$ is defined in the usual way.
For the off-shell lines,
the same $\xi_{\sst \rm Ref}$ is used  in all diagrams
contributing to a given amplitude.

For practical applications the authors of \cite{CSW} have chosen
$\xi_{\sst \rm Ref}^{\dot a}$ in \eqref{ofsh}
to be equal to $\tilde{\lambda}^{\dot a}$
of one of the external legs of negative helicity, e.g. the first one,
\be \label{ofsh2}
\xi_{\sst \rm Ref}=\ \tilde{\lambda}_1^{\dot a}
\ .
\ee
This corresponds to
identifying the reference spinor with one of the kinematic variables of the
theory. The explicit dependence on the reference spinor $\xi_{\sst \rm Ref}^{\dot a}$
disappears and the
resulting expressions for all scalar diagrams in the CSW approach are the functions
only of the kinematic variables $\lambda_{i\, a}$ and $\tilde{\lambda}^{\dot a}_i.$
This means that the expressions for all individual diagrams automatically
appear to be Lorentz-invariant
(in the sense that they do not depend on an external spinor $\xi_{\sst \rm Ref}^{\dot a}$)
and also gauge-invariant (since the reference spinor corresponds to  the
axial gauge fixing $n_\mu A^\mu =0,$ where
$n_{a\dot a} = \xi_{{\sst \rm Ref}\, a}\xi_{{\sst \rm Ref}\, \dot a}$).

There is a price to pay for this invariance of the individual diagrams.
Equations~\eqref{ofsh},\eqref{ofsh2} lead to unphysical
singularities\footnote{Unphysical means that these singularities are not the
standard IR soft and collinear divergences in the amplitudes.}
which occur for the whole of phase space and 
which have to be cancelled between the individual diagrams. The result for the
total amplitude is, of course, free of these unphysical singularities, but their cancellation
and the retention of the finite part requires some work, see \cite{CSW}
and section 3.1 of \cite{GK}.

It will be important for the purposes of this paper to note that
these unphysical singularities are specific to the three-gluon MHV vertices
and, importantly, they do not occur in any of the MHV vertices involving a fermion
field \cite{GK}.
To see how these singularities arise in gluon vertices, consider a 3-point MHV vertex,
\EQ{
A_3(g_1^-,g_2^-,g_3^+)=\ \frac{\vev{1~2}^4}{\vev{1~2}\vev{2~3}\vev{3~1}}
=\ \frac{\vev{1~2}^3}{\vev{2~3}\vev{3~1}}
\ . \label{simpl1}}
This vertex exists only when one of the legs is off-shell. Take it to be
the $g_3^+$ leg. Then Eqs.~\eqref{ofsh}, \eqref{ofsh2} give
\EQ{
\lambda_{3\,a}=\ (p_1+p_2)\, \tilde{\lambda}_1^{\dot a}=\
-\lambda_{1\, a}\,[1~1] - \, \lambda_{2\, a}\,[2~1] =\
 - \, \lambda_{2\, a}\,[2~1] \ .
}
This implies that $\vev{2~3}=-\vev{2~2}[2~1]=0,$
and the denominator of \eqref{simpl1} vanishes.
This is precisely the singularity we are after.
If instead of the $g_3^+$ leg, one takes the
$g_2^-$ leg go off-shell, then, $\vev{2~3}=-\vev{3~3}[3~1]=0$ again.

Now consider a three-point MHV vertex involving two fermions and a gluon,
\EQ{
A_3(\Lambda_1^-,g_2^-,\Lambda_3^+)=\ \frac{\vev{2~1}^3\vev{2~3}}{\vev{1~2}\vev{2~3}\vev{3~1}}
=\ -\, \frac{\vev{2~1}^2}{\vev{3~1}}
\ . \label{simpl2}}
Choose the reference spinor to be as before, $ \tilde{\lambda}_1^{\dot a},$ and
take the second or the third leg off-shell. This again makes $\vev{2~3}=0$,
but now the factor of $\vev{2~3}$ is cancelled on the right hand side of \eqref{simpl2}.
Hence, the vertex \eqref{simpl2} is regular, and 
there are no unphysical singularities
in the amplitudes involving at least one negative helicity fermion when it's helicity
is chosen to be the reference spinor \cite{GK}.
One concludes that the difficulties with singularities at intermediate stages
of the calculation occur only in purely gluonic amplitudes. One way
to avoid these intermediate singularities is to choose an off-shell continuation
different from the CSW prescription \eqref{ofsh},\eqref{ofsh2}.

Very recently,
Kosower \cite{Kosower} used an off-shell continuation by projection of the
off-shell momentum with respect to an on-shell reference momentum
$q_{\sst \rm Ref}^\mu,$ to derive, for the first time,
an expression for a general NMHV amplitude
with three negative helicity gluons. The amplitude in \cite{Kosower} was
from the start free of unphysical divergences, however it required
a certain amount of spinor algebra to bring it into the form
independent of the reference momentum.

Here we will propose another simple method for finding all purely gluonic NMHV
amplitudes. Using $\cN=1$ supersymmetric Ward identities one can relate
purely gluonic amplitudes to a linear combination of amplitudes with
one fermion--antifermion pair. As explained above,
the latter are free of singularities and are
manifestly Lorentz-invariant.
These fermionic amplitudes will be calculated in the following section
using the CSW scalar graph approach with fermions \cite{GK}.

To derive supersymmetric Ward identities
\cite{Grisaru} we use the fact that,
supercharges $Q$ annihilate the vacuum,
and consider the following equation,
\EQ{
\langle [Q\, , \,
\Lambda^+_k \ldots g_{r_1}^- \ldots g_{r_2}^- \ldots g_{r_3}^- \ldots ]\rangle
\ = \ 0
\ , }
where dots indicate positive helicity gluons. In order to make
anticommuting spinor $Q$ to be a singlet entering a commutative
(rather than anticommutative) algebra
with all the fields we contract it with a commuting spinor $\eta$ and
multiply it by a Grassmann number $\theta$. This defines a
commuting singlet operator $Q(\eta).$
Following \cite{Dixon} we can write down the following susy algebra relations,
\SP{\label{susyward}
[Q(\eta) \, , \, \Lambda^{+}(k)] \ = \ - \theta \vev{\eta~k}\,g^+ (k) \ , \quad
[Q(\eta) \, , \, \Lambda^{-}(k)] \ = \ + \theta [\eta~k]\,g^- (k) \ , \\
[Q(\eta) \, , \, g^{-}(k)] \ = \ + \theta \vev{\eta~k}\,\Lambda^- (k) \ , \quad
[Q(\eta) \, , \, g^{+}(k)] \ = \ - \theta [\eta~k]\,\Lambda^+ (k) \ .
}
In what follows, the anticommuting parameter
$\theta$ will cancel from the relevant expressions for the amplitudes.
The arbitrary spinors  $\eta_a,$ $\eta_{\dot a},$  will be fixed below.
It then follows from \eqref{susyward} that
\SP{\label{restggg}
{\vev{\eta~k}}\, A_n (g_{r_1}^- , g_{r_2}^- , g_{r_3}^-) = \
{\vev{\eta~r_1}}\,
A_n (\Lambda_{k}^+, \Lambda_{r_1}^-, g_{r_2}^-, g_{r_3}^-)
+ \,
{\vev{\eta~r_2}}\,
A_n (\Lambda_{k}^+, g_{r_1}^-, \Lambda_{r_2}^-, g_{r_3}^-)\\
+ \,
{\vev{\eta~r_3}}\,
A_n (\Lambda_{k}^+, g_{r_1}^-, g_{r_2}^-, \Lambda_{r_3}^- ) \ .
}
After choosing $\eta$ to be one of the three $r_j$ we find from
\eqref{restggg} that the purely gluonic amplitude with three negative helicities
is given by a sum of two fermion-antifermion-gluon-gluon amplitudes.
Note that in the expressions above and in what follows, in $n$-point amplitudes
we show only the relevant particles, and suppress all the positive
helicity gluons $g^+$.

Remarkably, this approach works for any number of negative helicities,
and the NMHV amplitude with $h$ negative gluons is expressed via a simple
linear combination of $h-1$ NMHV amplitudes with one fermion-antifermion pair.

In  sections 3 and 4 we will evaluate NMHV amplitudes with fermions.
In particular,
in section 3 we will calculate the following three amplitudes,
\EQ{\label{nprocesses}
A_n(\L_{m_1}^-,g_{m_2}^-,g_{m_3}^-,\L_k^+)\ , \quad
A_n(\L_{m_1}^-,g_{m_2}^-,\L_k^+,g_{m_3}^-)\ , \quad
A_n(\L_{m_1}^-,\L_k^+,g_{m_2}^-,g_{m_3}^-) \ .
}
In terms of these, the purely gluonic amplitude of \eqref{restggg}
reads
\SP{\label{restnew}
A_n (g_{r_1}^- , g_{r_2}^- , g_{r_3}^-) = \
-\frac{\vev{\eta~r_1}}{\vev{\eta~k}}\
A_n(\L_{m_1}^-,g_{m_2}^-,g_{m_3}^-,\L_k^+)|_{{m_1}=r_1, m_2=r_2, m_3=r_3} \\
- \,
\frac{\vev{\eta~r_2}}{\vev{\eta~k}}\
A_n(\L_{m_1}^-,g_{m_2}^-,\L_k^+,g_{m_3}^-)|_{{m_1}=r_2, m_2=r_3, m_3=r_1} \\
- \,\frac{\vev{\eta~r_3}}{\vev{\eta~k}}\
A_n(\L_{m_1}^-,\L_k^+,g_{m_2}^-,g_{m_3}^-)|_{{m_1}=r_3, m_2=r_1, m_3=r_2}
 \ ,
}
 and $\eta$ can be chosen to be one of the three $m_j$ to further
 simplify this formula.

\section{NMHV (- - -) Amplitudes with Two Fermions}

We start with the case of one fermion-antifermion pair, $\Lambda^-$, $\Lambda^+$,
and an arbitrary number of gluons, $g$. The amplitude has a schematic form,
$A_n(\L_{m_1}^-,g_{m_2}^-,g_{m_3}^-,\L_k^+),$
and without loss of generality we can have ${m_1}<m_2<m_3.$
With these conventions,
there are three different classes of amplitudes depending on the position
of the $\Lambda_k^+$ fermion relative to $m_1, m_2, m_3$:
\AL{\label{processes1}
A_n(\L_{m_1}^-,g_{m_2}^-,g_{m_3}^-,\L_k^+)\ , \\
\label{processes2}
A_n(\L_{m_1}^-,g_{m_2}^-,\L_k^+,g_{m_3}^-)\ , \\
\label{processes3}
A_n(\L_{m_1}^-,\L_k^+,g_{m_2}^-,g_{m_3}^-) \ .
}
Each of these three amplitudes receives contributions from different
types of scalar diagrams in the CSW approach. In all of these scalar
diagrams there are precisely two MHV vertices connected to each other
by a single scalar propagator \cite{CSW}. We will always arrange these diagrams
in such a way that the MHV vertex on the left has a positive helicity on the
internal line, and the right vertex has a negative helicity. Then, there are three
choices one can make \cite{Kosower} for the pair of negative helicity
particles to enter external lines of the left vertex,
$({m_1},m_2),$ $(m_2,m_3),$ or $(m_3,{m_1}).$  In addition to this, each diagram
in $\cN \le 1$ theory
corresponds to either a gluon exchange, or a fermion exchange.

The diagrams contributing to the first process \eqref{processes1}
are drawn in Figure 1. There are three gluon exchange diagrams
for all three partitions $(m_2,m_3),$ $({m_1},m_2),$ $(m_3,{m_1}),$
and there is one fermion exchange diagram for the partition $({m_1},m_2).$

It is now straightforward, using the expressions for the MHV vertices
\eqref{mpng},\eqref{ndcls},
to write down an analytic expression
for the
first diagram of Figure 1:
\SP{ \label{mmmPPP}
A_n^{(1)} = {1 \over \prod_{l=1}^n\ \vev{l~l+1}}
\sum_{i={m_1}}^{m_2-1} \sum_{j=m_3}^{k-1}{-\vev{(i+1,j)~ {m_1}}^3 \ \vev{(i+1,j)~ k}
\over
\vev{i~(i+1,j)}\vev{(i+1,j)~j+1} }\
{ \vev{i~i+1} \vev{j~j+1}    \over q_{i+1,j}^2}  \\
\times {\vev{m_2~m_3}^4 \over \vev{(j+1,i)~i+1} \vev{j~(j+1,i)}}
 \ .}
\begin{figure}[t]
\label{fig1}
\psfrag{i+}{\large$i\,+$}
\psfrag{i+1+}{\large$(i+1)\,+$}
\psfrag{j+}{\large$j\,+$}
\psfrag{j+1+}{\large$(j+1)\,+$}
\psfrag{k+}{\large$k\,+$}
\psfrag{n+}{\large$n\,+$}
\psfrag{n+}{\large$n\,+$}
\psfrag{1-}{\large${m_1}\,-$}
\psfrag{2-}{\large$2\,-$}
\psfrag{m2-}{\large$m_2\,-$}
\psfrag{m3-}{\large$m_3\,-$}
\psfrag{4}{\large$4\,+$}
\psfrag{+}{\large$+$}
\psfrag{-}{\large$-$}
\begin{center}
{\scalebox{0.62}{
\includegraphics{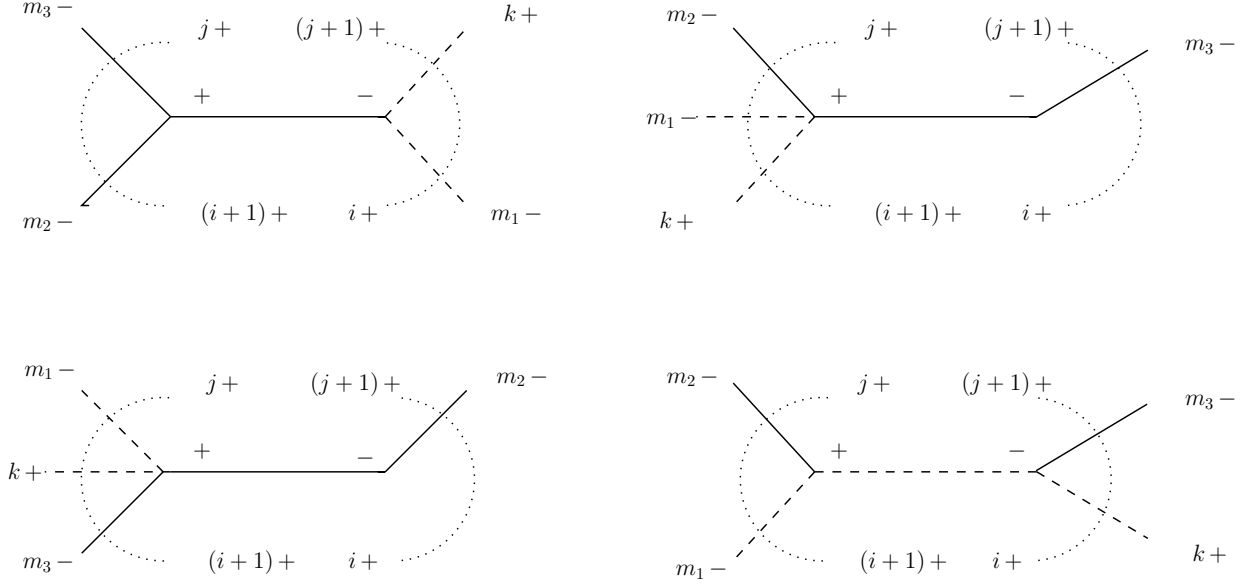}}
}
\end{center}
\caption{\it Tree diagrams with MHV vertices contributing to the
amplitude
$A_n(\L_{m_1}^-,g_{m_2}^-,g_{m_3}^-,\L_k^+)$.
Fermions, $\Lambda^+$ and $\Lambda^-,$ are represented by dashed lines and
negative helicity gluons, $g^-,$  by solid lines. Positive helicity gluons
$g^+$ emitted from each vertex are indicated by dotted semicircles with labels
showing the bounding $g^+$ lines in each MHV vertex. }
\end{figure}
This expression is a direct rendering of the `Feynman rules' for the
scalar graph method \cite{CSW,GK}, followed by factoring out the overall
factor of $(\prod_{l=1}^n\ \vev{l~l+1})^{-1}.$ The objects $(i+1,j)$ and
$(j+1,i)$ appearing on the right hand side of \eqref{mmmPPP}
denote the spinors $\lambda_{i+1,j}$ and $\lambda_{j+1,i}$
corresponding to the off-shell momentum $q_{i+1,j}$
\bea
&q_{i+1,j} \equiv\ p_{i+1} + p_{i+2}+ \ldots + p_j \ , \
q_{j+1,i} \equiv\ p_{j+1} + p_{j+2}+ \ldots + p_i \ , \
q_{i+1,j} + q_{j+1,i} =\ 0  \\
&\lambda_{i+1,j\,a} \equiv \ q_{i+1,j\, a\dot a}\, \xi_{\sst \rm Ref}^{\dot a} =\
- \lambda_{j+1,i\,a} \ ,
\eea
where $\xi_{\sst \rm Ref}^{\dot a}$ is the reference (dotted) spinor \cite{CSW}
as in Eq.~\eqref{ofsh}.
All other spinors $\lambda_i$ are on-shell and
$\vev{i~(j,k)}$ is an abbreviation for a spinor product
$\vev{\lambda_i,\lambda_{jk}}$.

Having the freedom to choose any reference spinor
we will always
choose it to be the spinor of the fermion $\L^-.$
In this section, this is the spinor of $\L_{m_1}^-,$
\be \label{ofshNEW}
\xi_{\sst \rm Ref}=\ \tilde{\lambda}_{m_1}^{\dot a}
\ .
\ee
We can now re-write
\SP{
&\vev{i~(i+1,j)}\vev{(i+1,j)~j+1}
\vev{(j+1,i)~i+1} \vev{j~(j+1,i)} \\
&=\vev{i^-|{q\!\!\!/}_{i+1,j}| m_1^-}
\vev{j+1^-|{q\!\!\!/}_{i+1,j}| m_1^-}
\vev{i+1^-|{q\!\!\!/}_{i+1,j}| m_1^-}
\vev{j^-|{q\!\!\!/}_{i+1,j}| m_1^-} \ ,}
and define a universal combination,
\EQ{ \label{Ddef}
D
=\vev{i^-|{q\!\!\!/}_{i+1,j}| m_1^-}
\vev{j+1^-|{q\!\!\!/}_{i+1,j}| m_1^-}
\vev{i+1^-|{q\!\!\!/}_{i+1,j}| m_1^-}
\vev{j^-|{q\!\!\!/}_{i+1,j}| m_1^-}\,
\frac{q_{i+1,j}^2}{\vev{i~i+1}\vev{j~j+1}}
}
Note that 
Here we introduced the standard Lorentz-invariant
matrix element
$\vev{ i^-|{{p\!\!\!/}\;}_k|j^-} =i^a \,p_{k~a\dot a}\,j^{\dot a}$,
which in terms of the spinor products  is
\be \label{lorinv}
\vev{ i^-|{{p\!\!\!/}\;}_k|j^-} = \
\vev{ i^-|^a \ |k^+}_a \ \vev{k^+|_{\dot a} \ |j^-}^{\dot a} = \
-\vev{i~k} \, [k~j] = \ \vev{i~k} \, [j~k] \ .
\ee

The expression for $A_n^{(1)} $ now becomes:
\SP{ \label{A1}
A_n^{(1)} = {-1 \over \prod_{l=1}^n\ \vev{l~l+1}}
\sum_{i=m_1}^{m_2-1} \sum_{j=m_3}^{k-1}{\vev{m_1^-|{q\!\!\!/}_{i+1,j}|m_1^-}^3
\vev{k^-|{q\!\!\!/}_{i+1,j}| m_1^-}\vev{m_2~m_3}^4
\over
D }\
\ .}

For the second diagram of Figure 1, we have
\SP{ \label{A2}
A_n^{(2)} = {-1 \over \prod_{l=1}^n\ \vev{l~l+1}}
\sum_{i=m_3}^{k-1} \sum_{j=m_2}^{m_3-1}{\vev{m_3^-|{q\!\!\!/}_{i+1,j}|m_1^-}^4
\vev{m_2~m_1}^3\vev{m_2~k}
\over
D }\
\ .}
The MHV vertex on the right
in the second diagram in Figure 1
can collapse to a 2-leg vertex. This occurs when $i=m_3$ and $j+1=m_3.$
This vertex is identically zero, since $q_{j+1,i}=p_{m_3}=-q_{i+1,j},$
and
$\vev{m_3~m_3}=0.$ Similar considerations apply in
\eqref{A3}, \eqref{A2'},
\eqref{A4'}, \eqref{A22},
\eqref{A23}, \eqref{A2''} and
\eqref{tA5}.

Expressions corresponding to the third and fourth diagrams in Figure 1 are
\begin{eqnarray}
 \label{A3}
A_n^{(3)} &=& {-1 \over \prod_{l=1}^n\ \vev{l~l+1}}
\sum_{i=m_2}^{m_3-1} \ \sum_{j=m_1}^{m_2-1}{\vev{m_2^-|{q\!\!\!/}_{i+1,j}|m_1^-}^4
\vev{m_3~m_1}^3\vev{m_3~k}
\over
D }\ , \\
\label{A4}
A_n^{(4)} &=& {-1 \over \prod_{l=1}^n\ \vev{l~l+1}}
\sum_{i=k}^{n+m_1-1} \ \sum_{j=m_2}^{m_3-1}{\vev{m_3^-|{q\!\!\!/}_{i+1,j}|m_1^-}^3
\vev{m_2^-|{q\!\!\!/}_{i+1,j}|m_1^-}
\vev{m_2~m_1}^3\vev{m_3~k}
\over
D } \ .
\end{eqnarray}
Note that the first sum in \eqref{A4}, $\sum_{i=k}^{n+m_1-1},$ is understood to run in cyclic
order, for example $\sum_{i=4}^{3} = \sum_{i=4,\ldots,n,1,2,3}.$ The same comment will also apply to
similar sums in
Eqs.~\eqref{A1'}, \eqref{A4'}, \eqref{A12}, \eqref{A22} below.

The total amplitude is the sum
of \eqref{A1}, \eqref{A2}, \eqref{A3} and \eqref{A4},
\bea
A_n(\L_{m_1}^-,g_{m_2}^-,g_{m_3}^-,\L_k^+) = \ \sum_{i=1}^{4} A_n^{(i)} \ .
\eea

There are three sources of zeroes in the denominator combination $D$ defined in
\eqref{Ddef}.   First, there are genuine zeroes in, for example,
$\vev{i^-|{q\!\!\!/}_{i+1,j}| m_1^-}$ when $q_{i+1,j}$ is proportional to
$p_i$.  This occurs when $j=i-1$.   Such terms are always associated with
two-leg vertices as discussed above and produce zeroes
in the numerator.   In fact, the number of zeroes in the numerator always
exceeds the number of zeroes in the denominator and this contribution vanishes.
Second, there are zeroes associated with three-point vertices when,  for example, $i=m_2$ and
$q_{i+1,j}= p_{m_2}+p_{m_1}$ so that 
$\vev{m_2^-|{m\!\!\!/}_{1}+{m\!\!\!/}_{2}| m_1^-} = 0$.
As discussed in Sec.~2, there is always a compensating factor in the numerator.
Such terms give a finite contribution (see \eqref{simpl2}).
Third, there are accidental zeroes when $q_{i+1,j}$ happens to be 
a linear combination of $p_i$ and $p_{m_1}$.   For general phase space points
this is not the case.
However, at certain phase space points,  the Gram determinant of 
$p_i$, $p_{m_1}$ and $q_{i+1,j}$ does vanish.  This produces an apparent
singularity in individual terms in \eqref{A1}--\eqref{A4} which   
cancels when all contributions are taken into account.
This cancellation can be achieved numerically or 
straightforwardly eliminated using standard spinor techniques~\cite{Kosower}.

For the special case of coincident negative helicities, $m_1=1$, $m_2 = 2$, $m_3 = 3$,
the double sums in Eqs.~\eqref{A1}--\eqref{A4} collapse to single sums.
Furthermore,
we see that the contribution from \eqref{A3} vanishes due to momentum
conservation, $q_{2,1}=0$.  The remaining three
terms agree with the result presented in Eq.~(3.6) of Ref.~\cite{GK}.

\newpage
We now consider the second amplitude,
Eq.~\eqref{processes2}. The scalar graph diagrams
are shown in Figure 2. There is a fermion exchange
and a gluon exchange diagram for two of the line assignments,
 $(m_1,m_2),$ and $(m_3,m_1),$  and none for the remaining assignment $(m_2,m_3).$
\begin{figure}[t]
\label{fig2}
\psfrag{i+}{\large$i\,+$}
\psfrag{i+1+}{\large$(i+1)\,+$}
\psfrag{j+}{\large$j\,+$}
\psfrag{j+1+}{\large$(j+1)\,+$}
\psfrag{k+}{\large$k\,+$}
\psfrag{n+}{\large$n\,+$}
\psfrag{n+}{\large$n\,+$}
\psfrag{1-}{\large$m_1\,-$}
\psfrag{2-}{\large$2\,-$}
\psfrag{m2-}{\large$m_2\,-$}
\psfrag{m3-}{\large$m_3\,-$}
\psfrag{+}{\large$+$}
\psfrag{-}{\large$-$}
\begin{center}
{\scalebox{0.62}{
\includegraphics{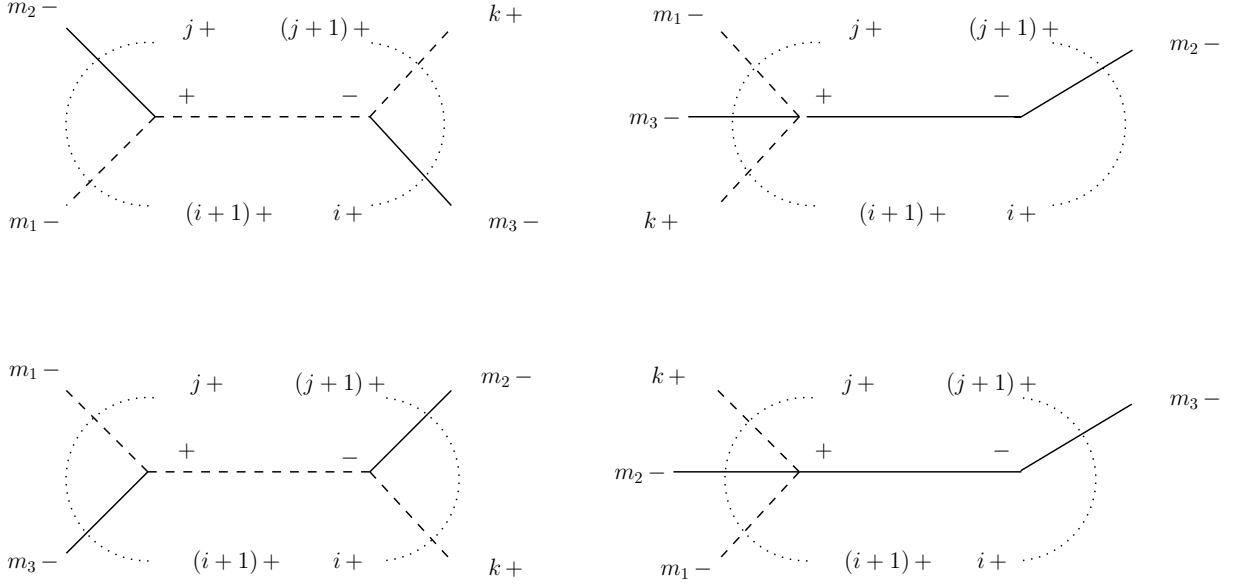}}
}
\end{center}
\caption{\it Tree diagrams with MHV vertices contributing to the
amplitude
$A_n(\L_{m_1}^-,g_{m_2}^-,\L_k^+,g_{m_3}^-)$.
}
\end{figure}

These four diagrams result in:
\begin{eqnarray}
 \label{A1'}
A_n^{(1)'} &=& {1 \over \prod_{l=1}^{n}\ \vev{l~l+1}}
\sum_{i=m_3}^{n+m_1-1} \ \sum_{j=m_2}^{k-1}{\vev{m_3^-|{q\!\!\!/}_{i+1,j}|m_1^-}^3
\vev{m_2^-|{q\!\!\!/}_{i+1,j}|m_1^-}
\vev{m_2~m_1}^3\vev{m_3~k}
\over
D }
 \\
 \label{A2'}
A_n^{(2)'} &=& {1 \over \prod_{l=1}^n\ \vev{l~l+1}}
\sum_{i=m_2}^{k-1}\ \sum_{j=m_1}^{m_2-1}{\vev{m_2^-|{q\!\!\!/}_{i+1,j}|m_1^-}^4
\vev{m_3~m_1}^3\vev{m_3~k}
\over
D }\
\ , \\
 \label{A3'}
A_n^{(3)'} &=& {1 \over \prod_{l=1}^n\ \vev{l~l+1}}
\sum_{i=k}^{m_3-1} \ \sum_{j=m_1}^{m_2-1}{\vev{m_2^-|{q\!\!\!/}_{i+1,j}|m_1^-}^3
\vev{m_3^-|{q\!\!\!/}_{i+1,j}|m_1^-}
\vev{m_3~m_1}^3\vev{m_2~k}
\over
D }
\ ,\\
 \label{A4'}
A_n^{(4)'} &=& {-1 \over \prod_{l=1}^n\ \vev{l~l+1}}
\sum_{i=m_3}^{n+m_1-1}\ \sum_{j=k}^{m_3-1}{\vev{m_3^-|{q\!\!\!/}_{i+1,j}|m_1^-}^4
\vev{m_2~m_1}^3\vev{m_2~k}
\over
D }\
\ ,
\end{eqnarray}
and the final answer for \eqref{processes2} is,
\bea
A_n(\L_{m_1}^-,g_{m_2}^-,\L_k^+,g_{m_3}^-) = \ \sum_{i=1}^{4}A_n^{(i)'} \ .
\eea

\newpage
Finally, we give the result for \eqref{processes3}. The corresponding diagrams
are drawn in Figure 3.
We find
\begin{eqnarray}
\label{A12}
A_n^{(1)''} &=& {1 \over \prod_{l=1}^n\ \vev{l~l+1}}
\sum_{i=k}^{m_2-1} \sum_{j=m_3}^{n+m_1-1}{\vev{m_1^-|{q\!\!\!/}_{i+1,j}|m_1^-}^3
\vev{k^-|{q\!\!\!/}_{i+1,j}|m_1^-}\vev{m_2~m_3}^4
\over
D }\
\ ,\\
 \label{A22}
A_n^{(2)''} &=& {1 \over \prod_{l=1}^n\ \vev{l~l+1}}
\sum_{i=m_3}^{n+m_1-1} \sum_{j=m_2}^{m_3-1}{\vev{m_3^-|{q\!\!\!/}_{i+1,j}|m_1^-}^4
\vev{m_2~m_1}^3\vev{m_2~k}
\over
D }\
\ ,\\
 \label{A23}
A_n^{(3)''} &=& {1 \over \prod_{l=1}^n\ \vev{l~l+1}}
\sum_{i=m_2}^{m_3-1} \sum_{j=k}^{m_2-1}{\vev{m_2^-|{q\!\!\!/}_{i+1,j}|m_1^-}^4
\vev{m_3~m_1}^3\vev{m_3~k}
\over
D }\ ,\\
 \label{A24}
A_n^{(4)''} &=& {-1 \over \prod_{l=1}^n\ \vev{l~l+1}}
\sum_{i=m_2}^{m_3-1} \sum_{j=m_1}^{k-1}{\vev{m_2^-|{q\!\!\!/}_{i+1,j}|m_1^-}^3
\vev{m_3^-|{q\!\!\!/}_{i+1,j}|m_1^-}
\vev{m_3~m_1}^3\vev{m_2~k}
\over
D }
\ .
\end{eqnarray}
As before, the full amplitude is given by the sum of contributions,
\bea
A_n(\L_{m_1}^-,\L_k^+,g_{m_2}^-,g_{m_3}^-) = \ \sum_{i=1}^{4}A_n^{(i)''} \ .
\eea
\begin{figure}[t]
\label{fig3}
\psfrag{i+}{\large$i\,+$}
\psfrag{i+1+}{\large$(i+1)\,+$}
\psfrag{j+}{\large$j\,+$}
\psfrag{j+1+}{\large$(j+1)\,+$}
\psfrag{k+}{\large$k\,+$}
\psfrag{n+}{\large$n\,+$}
\psfrag{n+}{\large$n\,+$}
\psfrag{1-}{\large$m_1\,-$}
\psfrag{2-}{\large$2\,-$}
\psfrag{m2-}{\large$m_2\,-$}
\psfrag{m3-}{\large$m_3\,-$}
\psfrag{+}{\large$+$}
\psfrag{-}{\large$-$}
\begin{center}
{\scalebox{0.62}{
\includegraphics{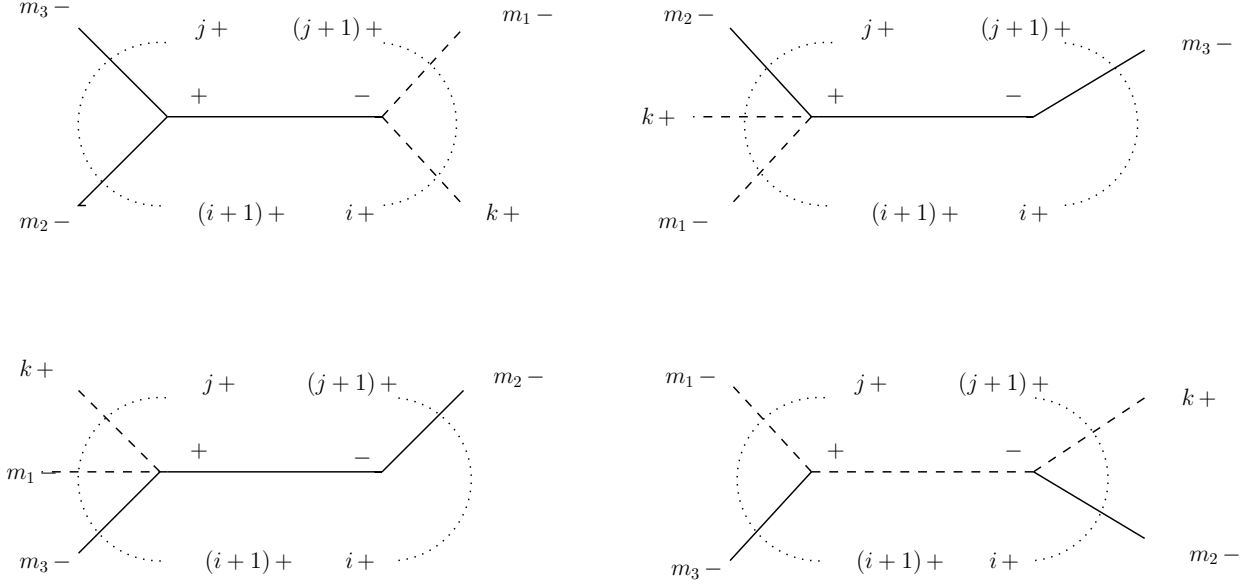}}
}
\end{center}
\caption{\it Tree diagrams with MHV vertices contributing to the
amplitude
$A_n(\L_{m_1}^-,\L_k^+,g_{m_2}^-,g_{m_3}^-)$ .
}
\end{figure}

\newpage

\section{NMHV (- - -) Amplitudes with Four  Fermions}

We now consider the amplitudes with 2 fermion-antifermion lines.
In what follows, without loss of generality we will choose the negative helicity gluon to be the
first particle. With this convention, we can write the six
inequivalent amplitudes as:
\AL{ \label{proc1}
A_n(g_1^-,\L_{m_2}^-,\L_{m_3}^-,\L_{m_p}^+,\L_{m_q}^+)\ , \\
\label{proc2}
A_n(g_1^-,\L_{m_2}^-,\L_{m_p}^+,\L_{m_3}^-,\L_{m_q}^+)\ ,\\
\label{proc3}
A_n(g_1^-,\L_{m_2}^-,\L_{m_p}^+,\L_{m_q}^+,\L_{m_3}^-)\ ,\\
\label{proc4}
A_n(g_1^-,\L_{m_p}^+,\L_{m_2}^-,\L_{m_3}^-,\L_{m_q}^+)\ ,\\
\label{proc5}
A_n(g_1^-,\L_{m_p}^+,\L_{m_2}^-,\L_{m_q}^+,\L_{m_3}^-)\ ,\\
\label{proc6}
A_n(g_1^-,\L_{m_p}^+,\L_{m_q}^+,\L_{m_2}^-,\L_{m_3}^-)\ .
}
\begin{figure}[b]
\label{fig4}
\psfrag{i+}{\large$i\,+$}
\psfrag{i+1+}{\large$(i+1)\,+$}
\psfrag{j+}{\large$j\,+$}
\psfrag{j+1+}{\large$(j+1)\,+$}
\psfrag{p+}{\large$p\,+$}
\psfrag{q+}{\large$q\,+$}
\psfrag{n+}{\large$n\,+$}
\psfrag{n+}{\large$n\,+$}
\psfrag{1-}{\large$1\,-$}
\psfrag{2-}{\large$2\,-$}
\psfrag{m2-}{\large$m_2\,-$}
\psfrag{m3-}{\large$m_3\,-$}
\psfrag{+}{\large$+$}
\psfrag{-}{\large$-$}
\begin{center}
{\scalebox{0.62}{
\includegraphics{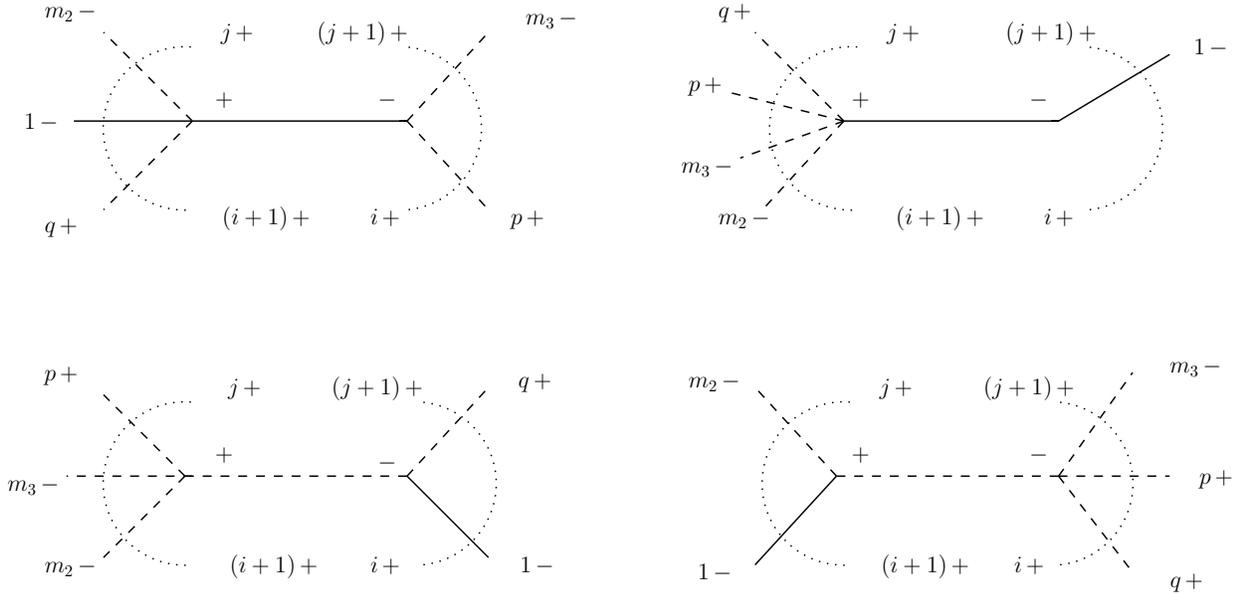}}
}
\end{center}
\caption{\it Tree diagrams with MHV vertices contributing to the
four fermion amplitude
$A_n(g_1^-,\L_{m_2}^-,\L_{m_3}^-,\L_{m_p}^+,\L_{m_q}^+)$.
}
\end{figure}

The calculation of the amplitudes of \eqref{proc1}-\eqref{proc6} is straightforward.
The diagrams contributing to the first process are shown in Figure~4.
It should be noted that not all the amplitudes in \eqref{proc1}-\eqref{proc6} receive
contributions from the same number of diagrams. For example, there are
four diagrams   for the  process of \eqref{proc1} while there are
six for that of \eqref{proc2}.
In order to avoid vanishing denominators, one can choose
the reference spinor to be $\tilde{\eta}=\tilde{\l}_{m_2}$.
With this choice the result can be written as:
\begin{eqnarray}
 \label{A1''}
\tilde{A}_n^{(1)} &=& {1 \over \prod_{l=1}^n\ \vev{l~l+1}}
\sum_{i=p}^{q-1} \sum_{j=m_2}^{m_3-1}{\vev{m_3^-|{q\!\!\!/}_{i+1,j}|m_2^-}^3
\vev{p^-|{q\!\!\!/}_{i+1,j}|m_2^-}
\vev{1~m_2}^3\vev{1~q}
\over
D }\
\ ,\\
\label{A2''}
\tilde{A}_n^{(2)} &=& {-1 \over \prod_{l=1}^n\ \vev{l~l+1}}
\sum_{i=1}^{m_2-1} \sum_{j=q}^{n}{\vev{1^-|{q\!\!\!/}_{i+1,j}|m_2^-}^4
\vev{m_2~m_3}^3\vev{p~q}
\over
D }\
 \ ,\\
 \label{A3''}
\tilde{A}_n^{(3)} &=& {1 \over \prod_{l=1}^n\ \vev{l~l+1}}
\sum_{i=1}^{m_2-1} \sum_{j=p}^{q-1}{\vev{1^-|{q\!\!\!/}_{i+1,j}|m_2^-}^3
\vev{p^-|{q\!\!\!/}_{i+1,j}|m_2^-}
\vev{m2~m_3}^3\vev{1~q}
\over
D }\
\ ,\\
\label{A4''}
\tilde{A}_n^{(4)} &=& {-1 \over \prod_{l=1}^n\ \vev{l~l+1}}
\sum_{i=q}^{n} \sum_{j=m_2}^{m_3-1}{\vev{m_3^-|{q\!\!\!/}_{i+1,j}|m_2^-}^3
\vev{1^-|{q\!\!\!/}_{i+1,j}|m_2^-}
\vev{1~m_2}^3\vev{p~q}
\over
D }\
\ .
\end{eqnarray}
As before the final result is the sum of Eq.~\eqref{A1''}-\eqref{A4''}.
\begin{equation}
A_n(g_1^-,\L_{m_2}^-,\L_{m_3}^-,\L_{m_p}^+,\L_{m_q}^+)=
\sum_{i=1}^4 \tilde{A}_n^{(i)}\ .
\end{equation}
Once again, for the case of coincident negative helicities, $m_2 = 2$, $m_3 =
3$, the double sums collapse to single summations and we recover
the results given in Ref.~\cite{GK}.

\begin{figure}[b]
\label{fig5}
\psfrag{i+}{\large$i\,+$}
\psfrag{i+1+}{\large$(i+1)\,+$}
\psfrag{j+}{\large$j\,+$}
\psfrag{j+1+}{\large$(j+1)\,+$}
\psfrag{p+}{\large$p\,+$}
\psfrag{q+}{\large$q\,+$}
\psfrag{n+}{\large$n\,+$}
\psfrag{n+}{\large$n\,+$}
\psfrag{1-}{\large$1\,-$}
\psfrag{2-}{\large$2\,-$}
\psfrag{m2-}{\large$m_2\,-$}
\psfrag{m3-}{\large$m_3\,-$}
\psfrag{+}{\large$+$}
\psfrag{-}{\large$-$}
\begin{center}
{\scalebox{0.42}{
\includegraphics{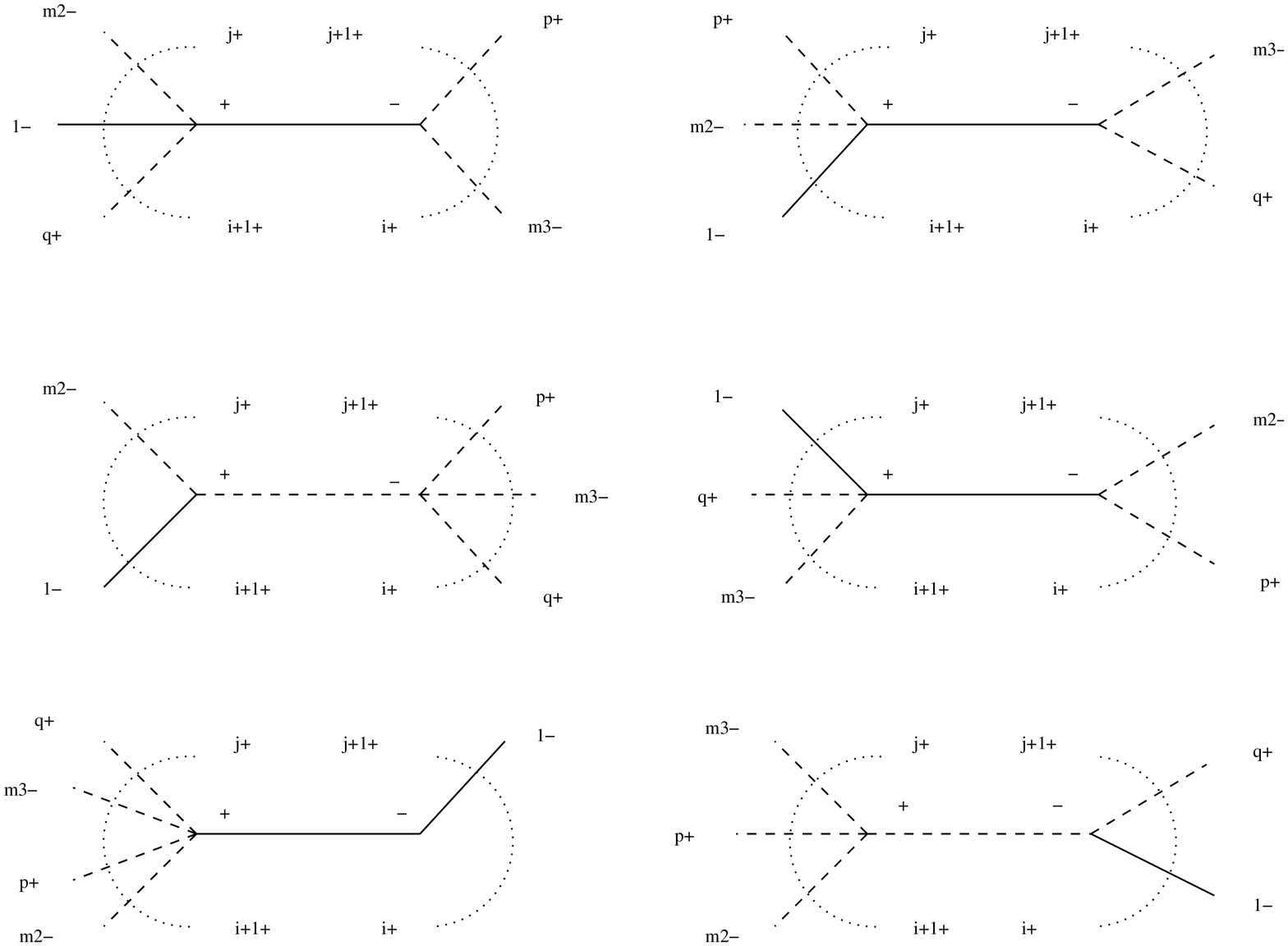}}
}
\end{center}
\caption{\it Tree diagrams with MHV vertices contributing to the
four fermion amplitude $A_n(g_1^-,\L_{m_2}^-,\L_{m_p}^+,\L_{m_3}^-,\L_{m_q}^+)$.
}
\end{figure}
As a last example we write down the expression for the amplitude
of \eqref{proc2}. The corresponding diagrams are shown in Figure 5.
We find,
\begin{eqnarray}
 \label{tA1}
\tilde{A}_n^{(1)'} &=& {-1 \over \prod_{l=1}^n\ \vev{l~l+1}}
\sum_{i=m_3}^{q-1} \sum_{j=m_2}^{p-1}{\vev{m_3^-|{q\!\!\!/}_{i+1,j}|m_2^-}^3
\vev{p^-|{q\!\!\!/}_{i+1,j}|m_2^-}
\vev{1~m_2}^3\vev{1~q}
\over
D }\
\ ,\\
\label{tA2}
\tilde{A}_n^{(2)'} &=& {1 \over \prod_{l=1}^n\ \vev{l~l+1}}
\sum_{i=q}^{n} \sum_{j=p}^{m_3-1}{\vev{m_3^-|{q\!\!\!/}_{i+1,j}|m_2^-}^3
\vev{q^-|{q\!\!\!/}_{i+1,j}|m_2^-}
\vev{1~m_2}^3\vev{1~p}
\over
D }\
\ ,\\
\label{tA3}
\tilde{A}_n^{(3)'} &=& {1 \over \prod_{l=1}^n\ \vev{l~l+1}}
\sum_{i=q}^{n} \sum_{j=m_2}^{p-1}{\vev{m_3^-|{q\!\!\!/}_{i+1,j}|m_2^-}^3
\vev{1^-|{q\!\!\!/}_{i+1,j}|m_2^-}
\vev{1~m_2}^3\vev{p~q}
\over
D }\
\ ,\\
\label{tA4}
\tilde{A}_n^{(4)'} &=& {1 \over \prod_{l=1}^n\ \vev{l~l+1}}
\sum_{i=p}^{m_3-1} \sum_{j=1}^{m_2-1}{\vev{m_2^-|{q\!\!\!/}_{i+1,j}|m_2^-}^3
\vev{p^-|{q\!\!\!/}_{i+1,j}|m_2^-}
\vev{1~m_3}^3\vev{1~q}
\over
D }\
\ ,\\
\label{tA5}
\tilde{A}_n^{(5)'} &=& {1 \over \prod_{l=1}^n\ \vev{l~l+1}}
\sum_{i=1}^{m_2-1} \sum_{j=q}^{n}{\vev{1^-|{q\!\!\!/}_{i+1,j}|m_2^-}^4
\vev{m_2~m_3}^3\vev{p~q}
\over
D }\
\ ,\\
\label{tA6}
\tilde{A}_n^{(6)'} &=& {-1 \over \prod_{l=1}^n\ \vev{l~l+1}}
\sum_{i=1}^{m_2-1} \sum_{j=m_3}^{q-1}{\vev{1^-|{q\!\!\!/}_{i+1,j}|m_2^-}^3
\vev{p^-|{q\!\!\!/}_{i+1,j}|m_2^-}
\vev{m_2~m_3}^3\vev{1~q}
\over
D }\
\ .\end{eqnarray}
The full amplitude is the sum of Eq.~\eqref{A1''}-\eqref{A4''}.
\begin{equation}
A_n(g_1^-,\L_{m_2}^-,\L_{m_p}^+,\L_{m_3}^-,\L_{m_q}^+)=
\sum_{i=1}^6 \tilde{A}_n^{(i)'}\ .
\end{equation}

We close this section by listing the inequivalent
NMHV amplitudes with three fermion--antifermion pairs.
There are ten such amplitudes since choosing the first particle
to be a negative helicity fermion we are left with five fermions
(two of which have negative helicity and three positive) which
should be distributed in all possible ways among themselves, and, in
addition there are $(n-6)$ positive helicity gluons.
Thus the number of different possible ways is $5!$.
However, the order of  the particles of the same helicity
is immaterial (since one can always choose $m_2\leq m_3$
and  $m_p\leq m_q \leq m_r)$. This means that we have to divide
 $5!$ by  $3!$ (for the positive helicity fermions) and by  $2!$
(for the negative helicity fermions.)
Thus there are ten different  fermion amplitudes. These
are listed below:
\SP{
A_n(\L_1^-,\L_{m_2}^-,\L_{m_3}^-,\L_{m_p}^+,\L_{m_q}^+,\L_{m_r}^+)\ , \quad
A_n(\L_1^-,\L_{m_2}^-,\L_{m_p}^+,\L_{m_3}^-,\L_{m_q}^+,\L_{m_r}^+)\ ,\\
A_n(\L_1^-,\L_{m_2}^-,\L_{m_p}^+,\L_{m_q}^+,\L_{m_3}^-,\L_{m_r}^+)\ , \quad
A_n(\L_1^-,\L_{m_p}^+,\L_{m_2}^-,\L_{m_3}^-,\L_{m_q}^+,\L_{m_r}^+)\ ,\\
A_n(\L_1^-,\L_{m_p}^+,\L_{m_2}^-,\L_{m_q}^+,\L_{m_3}^-,\L_{m_r}^+)\ , \quad
A_n(\L_1^-,\L_{m_p}^+,\L_{m_q}^+,\L_{m_2}^-,\L_{m_3}^-,\L_{m_r}^+)\ ,  \\
A_n(\L_1^-,\L_{m_p}^+,\L_{m_q}^+,\L_{m_r}^+,\L_{m_2}^-,\L_{m_3}^-)\ ,  \quad
A_n(\L_1^-,\L_{m_p}^+,\L_{m_q}^+,\L_{m_2}^-,\L_{m_r}^+,\L_{m_3}^-)\ , \\
A_n(\L_1^-,\L_{m_p}^+,\L_{m_2}^-,\L_{m_q}^+,\L_{m_r}^+,\L_{m_3}^-)\ , \quad
A_n(\L_1^-,\L_{m_2}^-,\L_{m_p}^+,\L_{m_q}^+,\L_{m_r}^+,\L_{m_3}^-)\ .
}
These amplitudes also present no difficulty, and they can be evaluated in
the same manner as before.

\section{Iterations of the Analytic Supervertex}

\subsection{Analytic Supervertex}

So far we have encountered three types of MHV amplitudes
\eqref{mpng}, \eqref{ndcls} and \eqref{ndcls2}. The key feature
which distinguishes these amplitudes is the fact that they
depend only on $\vev{\lambda_i~\lambda_j}$
spinor products,
and not on $[\tilde\lambda_i~ \tilde\lambda_i].$ We will call
such amplitudes analytic.

All analytic amplitudes in generic $0\le \cN \le 4$
gauge theories can be combined into a single $\cN=4$ supersymmetric
expression
of Nair \cite{Nair},
\be
A_n^{\cN=4} =\
\delta^{(8)} \left(\sum_{i=1}^n \lambda_{i a}
\eta^A_i \right)\
{1 \over \prod_{i=1}^n \vev{i~i+1}} \ .
\label{nair}
\ee
Here $\eta^A_i$ are anticommuting variables and $A=1,2,3,4$.
The Grassmann-valued delta function is defined in the usual way,
\EQ{\label{delta8}
\delta^{(8)} \left(\sum_{i=1}^n \lambda_{i a}
\eta^A_i \right) \equiv \ \prod_{A=1}^4 \, \hf
\left(\sum_{i=1}^n \lambda_{i }^a \eta^A_i \right)
\left(\sum_{i=1}^n \lambda_{i a} \eta^A_i \right) \ ,
}
where we have inserted factors of $\hf$ for future convenience.
Taylor expanding \eqref{nair} in powers of $\eta_i$, one can identify
each term in the expansion with a particular tree-level analytic amplitude
in the $\cN=4$ theory.  $(\eta_i)^k$ for $k=0,\ldots,4$ is interpreted as
the $i^{\rm th}$ particle with helicity $h_i=1-{k\over 2}$.
This implies that helicities take values,
$\{1,{1\over 2},0,-{1\over 2},-1\},$ which precisely correspond to those of
the $\cN=4$ supermultiplet,
$\{g^-,\lambda^{-}_A,\phi^{AB},\Lambda^{A+},g^+\}.$

It is straightforward to write down a general rule \cite{GK}
for associating a power of
$\eta$ with all component fields in $\cN=4$,
\SP{ \label{nrules}
g^{-}_i\ \sim\ \eta_i^1 \eta_i^2 \eta_i^3 \eta_i^4 \ , \quad
&\phi^{AB}_i \ \sim\  \eta_i^A \eta_i^B \ , \qquad
\Lambda^{A+}_i \ \sim\  \eta_i^A \ , \qquad \qquad
g^{+}_i\ \sim\  1  \ , \\
\Lambda^{-}_{1} \ \sim\  -\,\eta_i^2 \eta_i^3 \eta_i^4 \ , \quad
&\Lambda^{-}_{2\,i} \ \sim\  -\,\eta_i^1 \eta_i^3 \eta_i^4 \ , \quad
\Lambda^{-}_{3\,i} \ \sim\  -\,\eta_i^1 \eta_i^2 \eta_i^4 \ ,
\quad
\Lambda^{-}_{4\,i} \ \sim\  -\,\eta_i^1 \eta_i^2 \eta_i^3 \ .
}
The first MHV amplitude \eqref{mpng} is derived from \eqref{nair}
by using the dictionary \eqref{nrules}
and by selecting
the $(\eta_r)^4 \ (\eta_s)^4$ term in \eqref{nair}.
The second amplitude \eqref{ndcls}
follows from the $(\eta_t)^4 (\eta_r)^3 (\eta_s)^1$ term in \eqref{nair};
and the third amplitude
\eqref{ndcls2} is an
$(\eta_r)^3 (\eta_s)^1 (\eta_p)^3 (\eta_q)^1$ term.

All amplitudes following from  \eqref{nair}
are analytic in the sense that they depend only on $\vev{\lambda_i~\lambda_j}$
spinor products,
and not on $[\tilde\lambda_i~ \tilde\lambda_i].$
There is a large number of such component amplitudes for an
extended susy Yang-Mills, and what is remarkable, not all of these
amplitudes are MHV. The analytic amplitudes of the $\cN=4$ SYM
obtained from \eqref{nair}, \eqref{nrules} are
\SP{\label{listanal}
&A_n(g^-, g^-) \ , \quad
A_n(g^-,\Lambda_A^-,\Lambda^{A+}) \ , \quad
A_n(\Lambda_A^-,\Lambda_B^-,\Lambda^{A+},\Lambda^{B+}) \ , \\
&A_n(g^-,\Lambda^{1+},\Lambda^{2+},\Lambda^{3+},\Lambda^{4+} ) \ , \quad
A_n(\Lambda_A^-,\Lambda^{A+},\Lambda^{1+},\Lambda^{2+},\Lambda^{3+},\Lambda^{4+})
 \ ,   \\
&A_n(\Lambda^{1+},\Lambda^{2+},\Lambda^{3+},\Lambda^{4+},
\Lambda^{1+},\Lambda^{2+},\Lambda^{3+},\Lambda^{4+} ) \ , \quad
A_n(\overline{\phi}_{AB},\Lambda^{A+},\Lambda^{B+},\Lambda^{1+},\Lambda^{2+},\Lambda^{3+},\Lambda^{4+})
 \ ,  \\
&A_n(g^-,\overline{\phi}_{AB}, \phi^{AB}) \ , \quad
A_n(g^-,\overline{\phi}_{AB}, \Lambda^{A+},\Lambda^{B+}) \ , \quad
A_n(\Lambda_A^-,\Lambda_B^-, \phi^{AB}) \ , \\
&A_n(\Lambda_A^-,{\phi}^{AB},\overline{\phi}_{BC},\Lambda^{C+}) \ , \quad
A_n(\Lambda_A^-,\overline{\phi}_{AB},\Lambda^{A+}, \Lambda^{B+},\Lambda^{C+})
 \ , \\
&A_n( \overline{\phi}, {\phi}, \overline{\phi}, {\phi}) \ , \quad
A_n( \overline{\phi}, {\phi}, \overline{\phi},\Lambda^{+}, \Lambda^{+}) \ , \quad
A_n( \overline{\phi},\overline{\phi},\Lambda^{+}, \Lambda^{+},
\Lambda^{+}, \Lambda^{+}) \ ,
 }
where it is understood that
$\overline{\phi}_{AB}=\hf \epsilon_{ABCD}\phi^{CD}.$
In Eqs.~\eqref{listanal} we do not distinguish between the different particle orderings in the
amplitudes. The labels refer to supersymmetry multiplets, $A,B=1,\ldots,4.$
Analytic amplitudes in \eqref{listanal} include the familiar MHV amplitudes,
 \eqref{mpng}, \eqref{ndcls}, \eqref{ndcls2}, as well as more complicated classes
of amplitudes with external gluinos $\Lambda^A,$ $\Lambda^{B\neq A},$ etc, and
with external scalar fields $\phi^{AB}.$

The second and third lines in \eqref{listanal} are not even MHV
amplitudes, they have less than two negative helicities, and nevertheless,
these amplitudes are non-vanishing  in $\cN=4$ SYM.

All the analytic amplitudes listed in \eqref{listanal} can be calculated
directly from \eqref{nair}, \eqref{nrules}. There is a simple algorithm
for doing this.
\begin{enumerate}
\item{} For each amplitude in \eqref{listanal} substitute the
fields by their $\eta$-expressions \eqref{nrules}. There are precisely
eight $\eta$'s for each analytic amplitude.
\item{} Keeping track of the overall sign, rearrange the anticommuting
$\eta$'s into a product of four pairs:
$({\rm sign})\times
\eta^1_i \eta^1_j \,\eta^2_k \eta^2_l \,\eta^3_m \eta^3_n \,\eta^4_r \eta^4_s.$
\item{} The amplitude is obtained by replacing each pair $\eta^A_i \eta^A_j$
by the spinor product $\vev{i~j}$ and dividing by the usual denominator,
\EQ{\label{analsimp}
A_n = \ ({\rm sign})\times
\frac{\vev{i~j}\vev{k~l}\vev{m~n}\vev{r~s}}{\prod_{l=1}^n\ \vev{l~l+1}}
\ .
}
\end{enumerate}

\subsection{Scalar graphs with analytic vertices}

The conclusion we draw from the previous section is
that in the scalar graph formalism
in $\cN \le 4$ SYM, the
amplitudes are characterised not by a number of negative helicities,
but rather by the total number of $\eta$'s associated to each amplitude
via the rules \eqref{nrules}.

The vertices of the scalar graph method are the analytic vertices
\eqref{listanal} which are all of degree-8 in $\eta$. These vertices
are analytic (they depend only on $\vev{i~j}$ spinor products) and not
necessarily MHV. These are component vertices of a single analytic
supervertex\footnote{The list of component vertices \eqref{listanal} is obtained
by writing down all partitions of 8 into groups of 4, 3, 2 and 1.
For example, $A_n(g^-,\overline{\phi}_{AB}, \Lambda^{A+},\Lambda^{B+})$
follows from $8=4+2+1+1.$}
\eqref{nair}.
The analytic amplitudes are of degree-8 and they are the elementary
blocks of the scalar graph approach. The next-to-minimal case are
the amplitudes of degree-12 in $\eta$, and they are obtained by connecting
two analytic vertices \cite{Nair} with a scalar propagator $1/q^2.$
Each analytic vertex
contributes 8 $\eta$'s and a propagator removes 4. Scalar diagrams with
three degree-8 vertices give the degree-12 amplitude, etc.
In general,
all $n$-point amplitudes are characterised by a degree $8,12, 16, \ldots, (4n-8)$
which are obtained from scalar diagrams with $1,2,3, \ldots$ analytic
vertices.\footnote{In practice, one needs to know only the first half of these amplitudes,
since degree-$(4n-8)$ amplitudes are anti-analytic (formerly known as
googly $\overline{\rm MHV}$ and they are simply given by degree-$8^*$ amplitudes,
similarly degree-$(4n-12)$ are given by degree-$12^*$, etc.}
In the next section we derive a simple
expression for the first iteration of the degree-8 vertex. This iterative
process can be continued straightforwardly to higher orders.

\subsection{Two analytic supervertices}

\begin{figure}[ht]
\label{fig6}
\psfrag{i+1}{\large$(i+1)$}
\psfrag{i}{\large$i$}
\psfrag{j+1}{\large$(j+1)$}
\psfrag{j}{\large$j$}
\psfrag{n1}{\large$n_1$}
\psfrag{n2}{\large$n_2$}
\psfrag{n+}{\large$n\,+$}
\psfrag{ib}{\large$\bar{I}$}
\psfrag{I}{\large$I$}
\psfrag{2-}{\large$2\,-$}
\psfrag{m2-}{\large$m_2\,-$}
\psfrag{m3-}{\large$m_3\,-$}
\psfrag{+}{\large$+$}
\psfrag{-}{\large$-$}
\begin{center}
{\scalebox{0.5}{
\includegraphics{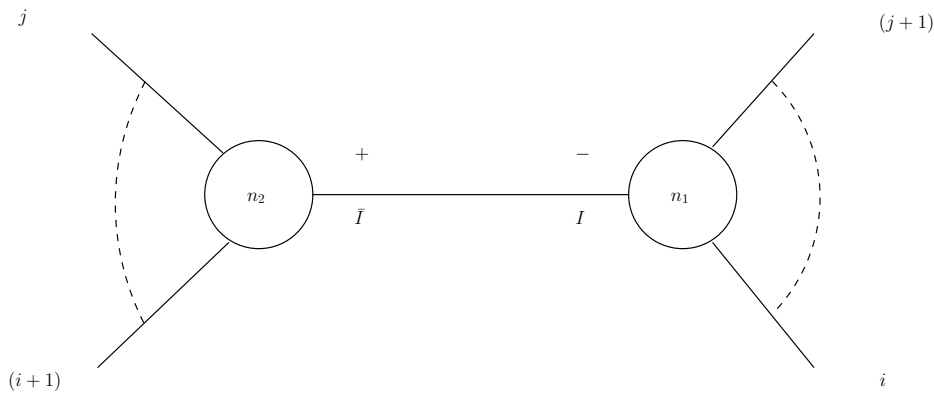}}
}
\end{center}
\caption{\it Tree diagrams with MHV vertices contributing to the
first amplitude of
Eq.~\eqref{proc2}.
}
\end{figure}
We now consider a diagram with two analytic supervertices
\eqref{nair} connected to one another by a single scalar propagator.
The diagram is depicted in Figure 6. We follow the same conventions
as in the previous sections, and the left vertex has a positive
helicity on the internal line $\bar{I}$, while the right vertex
has a negative helicity on the internal line $I$. The labelling of
the external lines in Figure 6 is also consistent with our conventions.
The right vertex has $n_1$ lines, and the left one has
$n_2$ lines in total, such that resulting amplitude $A_n$ has
$n=n_1+n_2-2$ external lines.
Suppressing summations over the distribution of $n_1$ and $n_2$
between the two vertices, we can write down an expression for the
corresponding amplitude which follows immediately from
\eqref{nair} and Figure 6:
\SP{ \label{itern1}
A_n =\ &{1 \over \prod_{l=1}^n\ \vev{l~l+1}}\
{1 \over q_I^2} \
{\vev{j~j+1} \vev{i~i+1} \over \vev{j~\bar{I}}\vev{\bar{I}~i+1}
\vev{i~I}\vev{I~j+1}} \\
&\times \int \prod_{A=1}^4 d \eta^A_I \
\delta^{(8)} \left(\lambda_{\bar{I}a}\eta_I^A +
\sum_{l_2 \neq \bar{I}}^{n_2} \lambda_{l_2 a}
\eta^A_{l_2} \right) \
\delta^{(8)} \left(\lambda_{Ia}\eta_I^A + \sum_{l_1 \neq I}^{n_1} \lambda_{l_1 a}
\eta^A_{l_1} \right)
 \ .
}
The two delta-functions in \eqref{itern1} come from the two
vertices \eqref{nair}. The summations in the delta-functions arguments run over the
$n_1-1$ external lines for right vertex, and $n_2-1$ external lines for the left one.
The integration over $d^4 \eta_I$ arises in \eqref{itern1} for the
following reason. Two separate (unconnected) vertices in Figure 6
would have $n_1+ n_2$ lines and, hence, $n_1+ n_2$ different
$\eta$'s (and $\lambda$'s). However the $I$ and the $\bar{I}$
lines are connected by the propagator, and there must be only
$n=n_1+n_2-2$ $\eta$-variables left. This is achieved in \eqref{itern1}
by setting
\EQ{ \eta_{\bar{I}}^A = \ \eta_{I}^A \ , }
and integrating over $d^4 \eta_I.$
The off-shell continuation of the internal spinors is defined
as before,
\EQ{
\lambda_{I a} = \ \sum_{l_1 \neq I}^{n_1}\,
p_{l_1\, a \dot{a}}\ \xi_{\sst\rm Ref}^{\dot a} = \ - \lambda_{\bar{I}a} \ .
\label{rlsls}}

We now integrate out four $\eta_I$'s which is made simple by rearranging the arguments of
the delta-functions via $\int \delta(f_2)\delta(f_1) =\int \delta(f_1+f_2)\delta(f_1),$
and noticing that the sum of two arguments, $f_1+f_2,$ does not depend on $\eta_I.$

The final result is
\EQ{\label{finrest}
A_n =\ {1 \over \prod_{l=1}^n\ \vev{l~l+1}}\
\delta^{(8)} \left(\sum_{i=1}^n \lambda_{i a} \eta^A_i \right)\
\prod_{A=1}^4\left(\sum_{l_1 \neq I}^{n_1}\, \vev{I~l_1} \eta^A_{l_1} \right) \
{1\over D} \ ,
}
and $D$ is the same as \eqref{Ddef} used in sections 3 and 4,
\EQ{\label{Pdef}
{1\over D}\ = \ {1 \over q_I^2} \
{\vev{j~j+1} \vev{i~i+1} \over \vev{j~I}\vev{I~i+1}
\vev{i~I}\vev{I~j+1}} \ .
}
There are 12 $\eta$'s in the superamplitude \eqref{finrest},
and the coefficients of the Taylor expansion in $\eta$'s give all the
component amplitudes of degree-12.

\section{Conclusions}

In this paper we have shown how all non-MHV tree-level amplitudes in $0\le \cN
\le 4$ gauge theories can be obtained directly from the known MHV amplitudes
using the scalar graph approach of Cachazo, Svrcek and Witten.  As a specific
example, we have focussed on amplitudes which are next-to-MHV, i.e. contain
three negative helicity particles and an arbitrary number of positive helicity
particles. By starting with amplitudes containing fermions, the reference
spinor for the  negative helicity gluons can be chosen to be that of the
negative helicity fermion.   As a consequence, the amplitudes are free of
unphysical singularities for generic phase space points
and no further helicity-spinor algebra is required to
convert the results into a numerically usable form. The gluons only amplitudes
can then be simply obtained as sums of fermionic amplitudes using the
supersymmetric Ward identity. These amplitudes are therefore also immediately
free of unphysical poles.  We have provided expressions for $(-,-,-)$
amplitudes with a two and four fermions and shown how to construct the
amplitudes for six fermions.   The extension to amplitudes with four or more
negative helicity particles is straightforward.
In principle one could use the results presented here to write a numerical
program for evaluating generic processes involving fermions and
bosons~\cite{LOrecur,LOdiagram}.

All of these results can be recovered from Nair's $\cN = 4$
supervertex.   This analytic vertex generates all possible interactions that
depend only on products of $\langle \lambda_i \lambda_j\rangle$.
Interestingly, all of the allowed vertices are not MHV.   For example,
$A_n(g^-,\Lambda^{1+},\Lambda^{2+},\Lambda^{3+},\Lambda^{4+} )$.   This implies
that the scalar graph approach is not primarily based on MHV
amplitudes.

The next logical step is to extend the formalism to the computation of loop graphs.
The twistor space
approach of  Ref.~\cite{Witten} may once again shed light on the structure of
gauge theory amplitudes at the loop level~\cite{CSW2}.
Also, the simplified (four-dimensional) helicity amplitudes for arbitrary
numbers of legs presented here may provide new impetus to computing loop
amplitudes in supersymmetric theories using the unitarity approach~\cite{BDK}
for sewing tree amplitudes to form loops.

\bigskip
\bigskip

\centerline{\bf Acknowledgements}

We thank Arnd Brandenburg, Lance Dixon and Gabriele 
Travaglini for illuminating discussions.
GG is supported by a grant from the State Scholarship Foundation
of Greece (I.K.Y.).
EWNG and VVK acknowledge PPARC Senior Fellowships.
This work was supported in part by
the EU Fifth Framework Programme  `Improving Human Potential', Research
Training Network `Particle Physics Phenomenology  at High Energy Colliders',
contract HPRN-CT-2000-00149.

\bigskip




\begin{thebibliography}{99}

\bibitem{CSW} F.~Cachazo, P.~Svrcek and E.~Witten,
``MHV vertices and tree amplitudes in gauge theory,''
hep-th/0403047.

\bibitem{Witten} E.~Witten,
``Perturbative gauge theory as a string theory in twistor space,''
hep-th/0312171.

\bibitem{GK}
G.~Georgiou and V.~V.~Khoze,
``Tree amplitudes in gauge theory as scalar MHV diagrams,''
JHEP {\bf 0405} (2004) 070,
hep-th/0404072.

\bibitem{Zhu}
C.~J.~Zhu,
``The googly amplitudes in gauge theory,''
JHEP {\bf 0404} (2004) 032
hep-th/0403115;\\
J.~B.~Wu and C.~J.~Zhu,
``MHV vertices and scattering amplitudes in gauge theory,''
hep-th/0406085.

\bibitem{BBK}
I.~Bena, Z.~Bern and D.~A.~Kosower,
``Twistor-space recursive formulation of gauge theory amplitudes,''
hep-th/0406133.

\bibitem{Kosower}
D.~A.~Kosower,
``Next-to-maximal helicity violating amplitudes in gauge theory,''
hep-th/0406175.

\bibitem{CSW2}
F.~Cachazo, P.~Svrcek and E.~Witten,
``Twistor space structure of one-loop amplitudes in gauge theory,''
hep-th/0406177.

\bibitem{BM}
N.~Berkovits,
``An alternative string theory in twistor space for N = 4 super-Yang-Mills,''
hep-th/0402045.
\\
N.~Berkovits and L.~Motl,
``Cubic twistorial string field theory,''
JHEP {\bf 0404} (2004) 056
hep-th/0403187.


\bibitem{RSV}
R.~Roiban, M.~Spradlin and A.~Volovich,
``A googly amplitude from the B-model in twistor space,''
JHEP {\bf 0404} (2004) 012
hep-th/0402016;
\\
R.~Roiban, M.~Spradlin and A.~Volovich,
``On the tree-level S-matrix of Yang-Mills theory,''
hep-th/0403190.


\bibitem{NV}
A.~Neitzke and C.~Vafa,
``N = 2 strings and the twistorial Calabi-Yau,''
hep-th/0402128.
\\
N.~Nekrasov, H.~Ooguri and C.~Vafa,
``S-duality and topological strings,''
hep-th/0403167.

\bibitem{W}
E.~Witten,
``Parity invariance for strings in twistor space,''
hep-th/0403199.

\bibitem{GLMN}
S.~Gukov, L.~Motl and A.~Neitzke,
``Equivalence of twistor prescriptions for super Yang-Mills,''
hep-th/0404085.

\bibitem{Siegel}
W.~Siegel,
``Untwisting the twistor superstring,''
hep-th/0404255.

\bibitem{Giombi}
S.~Giombi, R.~Ricci, D.~Robles-Llana and D.~Trancanelli,
``A note on twistor gravity amplitudes,''
hep-th/0405086.

\bibitem{Popov}
A.~D.~Popov and C.~Saemann,
 ``On supertwistors, the Penrose-Ward transform and N = 4 super Yang-Mills
theory,''
hep-th/0405123.

\bibitem{BW}
N.~Berkovits and E.~Witten,
``Conformal supergravity in twistor-string theory,''
hep-th/0406051.


\bibitem{Nair} V.~P.~Nair,
``A Current Algebra For Some Gauge Theory Amplitudes,''
Phys.\ Lett.\ B {\bf 214} (1988) 215.

\bibitem{Grisaru}
M.~T.~Grisaru, H.~N.~Pendleton and P.~van Nieuwenhuizen,
``Supergravity And The S Matrix,''
Phys.\ Rev.\ D {\bf 15} (1977) 996;
\\
M.~T.~Grisaru and H.~N.~Pendleton,
``Some Properties Of Scattering Amplitudes In Supersymmetric Theories,''
Nucl.\ Phys.\ B {\bf 124} (1977) 81.

\bibitem{PT} S.~J.~Parke and T.~R.~Taylor,
``An Amplitude For N Gluon Scattering,''
Phys.\ Rev.\ Lett.\  {\bf 56} (1986) 2459.

\bibitem{BG} F.~A.~Berends and W.~T.~Giele,
``Recursive Calculations For Processes With N Gluons,''
Nucl.\ Phys.\ B {\bf 306} (1988) 759.

\bibitem{Berends}
F.~A.~Berends, R.~Kleiss, P.~De Causmaecker, R.~Gastmans and T.~T.~Wu,
Phys.\ Lett.\ B {\bf 103} (1981) 124;
\\
P.~De Causmaecker, R.~Gastmans, W.~Troost and T.~T.~Wu,
Nucl.\ Phys.\ B {\bf 206} (1982) 53;
\\
R.~Kleiss and W.~J.~Stirling,
Nucl.\ Phys.\ B {\bf 262} (1985) 235;
\\
J.~F.~Gunion and Z.~Kunszt,
Phys.\ Lett.\ B {\bf 161} (1985) 333.

\bibitem{MP} M.~L.~Mangano and S.~J.~Parke,
``Multiparton Amplitudes In Gauge Theories,''
Phys.\ Rept.\  {\bf 200} (1991) 301.

\bibitem{Dixon} L.~J.~Dixon,
``Calculating scattering amplitudes efficiently,''
hep-ph/9601359.

\bibitem{LOrecur}
F.~A.~Berends, W.~T.~Giele and H.~Kuijf,
Phys.\ Lett.\ B {\bf 232}, 266 (1989);\\
F.~A.~Berends, H.~Kuijf, B.~Tausk and W.~T.~Giele,
Nucl.\ Phys.\ B {\bf 357}, 32 (1991); \\
F.~Caravaglios and M.~Moretti,
Phys.\ Lett.\ B {\bf 358}, 332 (1995)
hep-ph/9507237;\\
P.~Draggiotis, R.~H.~Kleiss and C.~G.~Papadopoulos,
Phys.\ Lett.\ B {\bf 439}, 157 (1998)
hep-ph/9807207;\\
P.~D.~Draggiotis, R.~H.~Kleiss and C.~G.~Papadopoulos,
Eur.\ Phys.\ J.\ C {\bf 24}, 447 (2002)
hep-ph/0202201;\\
M.~L.~Mangano, M.~Moretti, F.~Piccinini, R.~Pittau and A.~D.~Polosa,
JHEP {\bf 0307}, 001 (2003)
hep-ph/0206293.

\bibitem{LOdiagram}
T.~Stelzer and W.~F.~Long,
Comput.\ Phys.\ Commun.\  {\bf 81}, 357 (1994)
hep-ph/9401258;\\
A.~Pukhov {\it et al.},
hep-ph/9908288;\\
F.~Yuasa {\it et al.},
Prog.\ Theor.\ Phys.\ Suppl.\  {\bf 138}, 18 (2000)
hep-ph/0007053;\\
F.~Krauss, R.~Kuhn and G.~Soff,
JHEP {\bf 0202}, 044 (2002)
hep-ph/0109036;\\
F.~Maltoni and T.~Stelzer,
JHEP {\bf 0302} (2003) 027
hep-ph/0208156.


\bibitem{BDK}
Z.\ Bern, L.\ J.\ Dixon, D.\ C.\ Dunbar and D.\ A.\ Kosower,
``Fusing gauge theory tree amplitudes into loop amplitudes,''
Nucl.\ Phys.\ B435:59 (1995) hep-ph/9409265;\\
Z.\ Bern, L.\ J.\ Dixon and D.\ A.\ Kosower,
``Unitarity-based techniques for one-loop calculations in QCD,''
Nucl.\ Phys.\ Proc.\ Suppl.\  51C:243 (1996)
hep-ph/9606378.










\end{thebibliography}
\end{document}